\documentclass[aps,prb,twocolumn,amssymb,showpacs,amsmath,superscriptaddress,floatfix]{revtex4}
\usepackage{graphicx}
\usepackage{bm}

\begin{document}
\title {Superconducting proximity effect to the block antiferromagnetism in K$_{y}$Fe$_{2-x}$Se$_{2}$}
\author{Hong-Min Jiang}
\affiliation{Department of Physics and Center of Theoretical and
Computational Physics, The University of Hong Kong, Hong Kong,
China}
\affiliation{Department of Physics, Hangzhou Normal
University, Hangzhou, China}
\author{Wei-Qiang Chen}
\affiliation{Department of Physics, South University of Science and
Technology of China, Shenzhen, China}
\affiliation{Department of
Physics and Center of Theoretical and Computational Physics, The
University of Hong Kong, Hong Kong, China}
\author{Zi-Jian Yao}
\affiliation{Department of Physics and Center of Theoretical and
Computational Physics, The University of Hong Kong, Hong Kong,
China}
\author{Fu-Chun Zhang}
\affiliation{Department of Physics and Center of Theoretical and
Computational Physics, The University of Hong Kong, Hong Kong,
China}
\affiliation{Department of Physics, Zhejiang University,
Hangzhou, China}

\date{\today}

\begin{abstract}
Recent discovery of superconducting (SC) ternary iron selenides has
block antiferromagentic (AFM) long range order. Many experiments
show possible mesoscopic phase separation of the superconductivity
and antiferromagnetism, while the neutron experiment reveals a
sizable suppression of magnetic moment due to the superconductivity
indicating a possible phase coexistence. Here we propose that the
observed suppression of the magnetic moment may be explained due to
the proximity effect within a phase separation scenario. We use a
two-orbital model to study the proximity effect on a layer of block
AFM state induced by neighboring SC layers via an interlayer
tunneling mechanism. We argue that the proximity effect in ternary
Fe-selenides should be large because of the large interlayer
coupling and weak electron correlation. The result of our mean field
theory is compared with the neutron experiments semi-quantitatively.
The suppression of the magnetic moment due to the SC proximity
effect is found to be more pronounced in the $d$-wave
superconductivity and may be enhanced by the frustrated structure of
the block AFM state.
\end{abstract}

 \maketitle

\section{introduction}
The recent discovery of high-$T_{c}$ superconductivity in the
ternary iron selenides A$_{y}$Fe$_{2-x}$Se$_{2}$ (A=K; Rb;
Cs;...)~\cite{JGuo,AKrzton,MHFang} has triggered a new surge of
interest in study of iron-based superconductors (Fe-SC). The
fascinating aspect of these material lies in the tunable
Fe-vacancies in these materials, which substantially modifies the
normal-state metallic behavior and enhances the transition
temperature $T_{c}$ to above 30K from 9K for the binary system FeSe
at ambient pressure.~\cite{JGuo,MHFang,YZhang1} Particular attention
has been focused on the vacancy ordered 245 system,
K$_{0.8}$Fe$_{1.6}$Se$_{2}$, as it introduces a novel magnetic
structure into the already rich magnetism of Fe-SC. Unlike the
collinear~\cite{Cruz1,FMa1,XWYan1} or
bi-collinear~\cite{FMa2,WBao2,SLi1} AFM order observed in the parent
compounds of other Fe-SC, the neutron diffraction experiment has
clearly shown that these materials have a block AFM (BAFM)
order.~\cite{WBao1} Meanwhile, the AFM order with an unprecedentedly
large magnetic moment of $3.31\mu_{B}$/Fe below the N\'{e}el
temperature is the largest one among all the known parent compounds
of Fe-SC.~\cite{ZShermadini,WBao1} Moreover, the carrier
concentration is extremely low, indicating the parent compound to be
a magnetic insulator/semiconductor,~\cite{yzhou,RYu} in comparison
with a metallic spin-density-wave (SDW) state of the parent compound
in other Fe-SC.~\cite{RHYuan,MHFang}

The relation between the novel magnetism and superconductivity in
ternary Fe selenides is currently an interesting issue under debate.
The question is whether the superconductivity and the BAFM order are
phase separated or co-exist in certain region of the phase diagram.
The neutron experiment shows the suppression of the AFM ordering
below SC transition point,~\cite{WBao1} suggesting the coexistence.
Some other experiments, such as two-magnon
Raman-scattering,~~\cite{AMZhang} and muon-spin rotation and
relaxation ~\cite{Shermadini1} are consistent with this picture. On
the other hand, the ARPES,~\cite{FChen1} NMR ~\cite{Torchetti1} and
TEM ~\cite{ZWang1} experiments indicate a mesoscopic phase
separation between the superconductivity and the insulating AFM
state. Most recently, Li \textit{et al.} showed the
superconductivity and the BAFM orders to occur at different layers
of the Fe-selenide  planes in the STM measurement.~\cite{wli1}

The vacancy in Fe-selenides is an interesting but complicated issue.
The vacancy in the Fe-selenide carries a negative charge since the
Fe-ion has a valence of 2+. In the equilibrium, we expect the
vacancies to repel each other at short distance for the Coulomb
interaction and to attract to each other at a long distance for the
elastic strain. Such a scenario would be in favor of the phase
separation to form a vacancy rich and vacancy poor regions in the
compound. The challenge is then to explain the observed suppression
of magnetic moment of the BAFM due to the superconductivity. At the
phenomenological level, the suppression of magnetism due to
superconductivity has been reported previously,~\cite{aeppli} and
such phenomenon may be explained by Ginzburg-Landau
theory.~\cite{varma}

In this paper, we propose that the proximity effect of
superconductivity to the BAFM in a mesoscopically phase separated
Fe-selenides may be large to account the suppression of the AFM
moments observed in neutron experiment. More specifically, we use a
microscopic model to study the proximity effect on a layer of the
BAFM state induced by adjacent SC layer. The proximity effect in
Fe-selenides is expected to be important for the two reasons. One is
the weaker correlation effect, and the other is the larger
interlayer hopping amplitude, compared with those in cuprates. Both
of them may enhance proximity effect on the magnetism from the
neighboring SC layer. Our model calculations show the proximity
effect in a mesoscopically phase separated state of Fe-selenides may
explain various seemingly conflicted experiments.

\section{MODEL AND MEAN FIELD THEORY}
\begin{figure}[htb]
\begin{center}
\vspace{-.2cm}
\includegraphics[width=230pt,height=190pt]{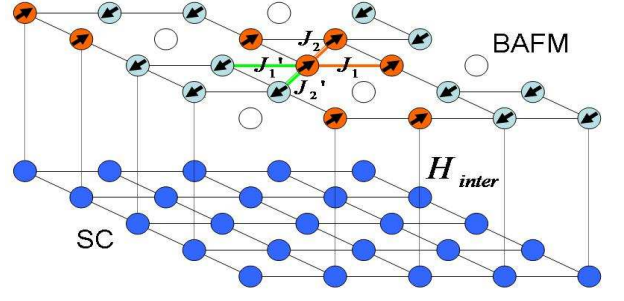}
\caption{(color online) Schematic diagram of the system in our model
to study proximity effect to the block AFM state (upper layer)
induced by superconductivity at the lower layer via a pair tunneling
process $H_{inter}$ in Eq. (3).}\label{fig1}
\end{center}
\end{figure}

A$_{y}$Fe$_{2-x}$Se$_{2}$ is a layered material with FeSe layers
separated by alkali atoms, similar to the 122 material in iron
pnictides family. To investigate the proximity effect to the BAFM
layer, we consider a single BAFM layer next to a SC layer as shown
schematically in Fig.~\ref{fig1}.  The electronic Hamiltonian
describing the BAFM layer is given by
\begin{align}
\label{eq:1}
H &=H_0 + H_{inter},
\end{align}
where $H_0$ describes the electron motion and spin couplings in the
BAFM layer and $H_{inter}$ describes the coupling to the neighboring
SC layer. We consider a two-orbital model to describe $H_0$,
\begin{eqnarray}
H_{0}=&&-\sum_{ij,\alpha\beta,\sigma}t_{ij,\alpha\beta}
C^{\dag}_{i,\alpha\sigma}C_{j,\beta\sigma}
-\mu\sum_{i,\alpha\sigma}C^{\dag}_{i,\alpha\sigma}C_{i,\alpha\sigma} \nonumber\\
&&+J_{1}\sum_{<ij>,\alpha\beta}\textbf{S}_{i,\alpha}\cdot\textbf{S}_{j,\beta}
+J_{2}\sum_{<<ij>>,\alpha\beta}\textbf{S}_{i,\alpha}\cdot\textbf{S}_{j,\beta}
\nonumber\\
&&+J'_{1}\sum_{<ij>',\alpha\beta}\textbf{S}_{i,\alpha}\cdot\textbf{S}_{j,\beta}
+J'_{2}\sum_{<<ij>>',\alpha\beta}\textbf{S}_{i,\alpha}\cdot\textbf{S}_{j,\beta},
\end{eqnarray}
where $C_{i,\alpha\sigma}$ annihilates an electron at site $i$ with
orbital $\alpha$ ($d_{xz}$ and $d_{yz}$) and spin $\sigma$, $\mu$ is
the chemical potential. $t_{ij,\alpha\beta}$ are the hopping
integrals, and $<ij>$ ($<ij>'$) and $<<ij>>$ ($<<ij>>'$) denote the
intra-block (inter-block) nearest (NN) and next nearest neighbor
(NNN) bonds, respectively [see the upper layer in Fig.~\ref{fig1}].
$J_1$ ($J_2$) are the exchange coupling constants for NN (NNN) spins
in the same block, and $J'_1$  ($J'_2$) are for the two NN (NNN)
spins in different blocks. The two-orbital model is a crude
approximation for electronic structure. However, it may be a minimal
model to capture some of basic physics in examining the proximity
effect. The band structure around the obtained Fermi energy from the
two-orbital model is shown in Fig.~\ref{fig2}, which is very similar
to the result obtained in density functional theory. The main
shortcoming in using the two-orbital model is that the magnetic
moment is $2\mu$B at largest, smaller than the experimentally
measured $3.31 \mu$B. We consider this to be a quantitative issue,
and will not qualitatively change our results.

We now consider $H_{inter}$, the coupling between SC layer and BAFM
layer. Because of the semiconducting gap of the BAFM layer and the
SC gap in the SC layer, the leading order of the interlayer coupling
are the pairing hopping between the SC and BAFM layers. According to
the crystal structures,~\cite{wli1} such a coupling term can be
expressed as ~\cite{mcmillan1}
\begin{eqnarray}
H_{inter}=&&\frac{t^2_{\tau}}{\omega_{c}}\sum_{ij,\alpha\beta,\alpha'\beta',\sigma}
(C^{\dag}_{i,\alpha,\sigma}C^{B}_{i,\beta,\sigma}C^{\dag}_{j,\alpha',\bar{\sigma}}C^{B}_{j,\beta',\bar{\sigma}}
\nonumber\\ &&+H.C.),
\end{eqnarray}
where $\omega_{c}$ is a characteristic energy, and $t_{\tau}$ is the
interlayer hopping integral, the superscript $B$ represents the SC
layer. With the mean field approximation
$\Delta_{ij,\alpha\alpha',\sigma\bar{\sigma}}=<C^{B}_{i,\alpha,\sigma}C^{B}_{j,\alpha',\bar{\sigma}}>$,
we have
\begin{eqnarray}
H_{inter}=\sum_{ij,\beta\beta',\sigma}(V_{\tau, ij}
C^{\dag}_{i,\beta,\sigma}C^{\dag}_{j,\beta',\bar{\sigma}}+H.C.),
\end{eqnarray}
where $V_{\tau, ij}= \frac{t^2_{\tau}}{\omega_{c}}
\sum_{\alpha\alpha'}\Delta_{ij,\alpha\alpha',\sigma\bar{\sigma}}$.

The $\sqrt{5}\times\sqrt{5}$ vacancy order and the BAFM order lead
to an enlarged unit cell with eight sites per unit cell. We use a
mean field theory for the Ising spins in Eq. (2) and obtain the
Bogoliubov-de Gennes equations in the enlarged unit cell
\begin{eqnarray}
\sum_{k}{'}\sum_{j,\beta}\left(
\begin{array}{lr}
H_{ij,\alpha\beta,\sigma}+\tilde{H}_{ij,\alpha\beta,\sigma} &
H_{c,ij,\alpha\beta} \\
H^{\ast}_{c,ij,\alpha\beta}&
-H^{\ast}_{ij,\alpha\beta,\bar{\sigma}}+\tilde{H}_{ij,\alpha\beta,\sigma}
\end{array}
\right) \nonumber\\
\times\exp[i\textbf{k}\cdot(\textbf{r}_{j}-\textbf{r}_{i})]\left(
\begin{array}{lr}
u^{k}_{n,j,\beta,\sigma} \\
v^{k}_{n,j,\beta,\bar{\sigma}}
\end{array}
\right)= E^{k}_{n}\left(
\begin{array}{lr}
u^{k}_{n,i,\alpha,\sigma} \\
v^{k}_{n,i,\alpha,\bar{\sigma}}
\end{array}
\right),
\end{eqnarray}
where, the summation of $k$ are over the reduced Brillouin zone,
and
\begin{eqnarray}
H_{ij,\alpha\beta,\sigma}=&&-t_{ij,\alpha\beta}-\mu,\nonumber\\
\tilde{H}_{ij,\alpha\beta,\sigma}=&&\sum_{\tau}(J_{\tau,intra}+J_{\tau,inter})<S_{i+\tau,\beta}>\delta_{ij}
\nonumber\\
H_{c,ij,\alpha\beta}=&&V_{\tau,ij}\sum_{\alpha'\beta'}
\Delta_{ij,\alpha'\beta',\sigma\bar{\sigma}}.
\end{eqnarray}
Here, $J_{\tau,intra}=J_{1}$ ($J_{\tau,inter}=J'_{1}$) if $\tau=\pm
\hat{x},\pm \hat{y}$ and $J_{\tau,intra}=J_{2}$
($J_{\tau,inter}=J'_{2}$) if $\tau=\pm \hat{x}\pm \hat{y}$.
$\hat{x}$ and $\hat{y}$ denote the unit vectors along $x$ and $y$
directions, respectively. $<S_{i+\tau,\beta}>$ is defined as
$(n_{i+\tau,\beta,\uparrow}-n_{i+\tau,\beta,\downarrow})/2$.
$u^{k}_{n,j,\alpha,\sigma}$ ($u^{k}_{n,j,\beta,\bar{\sigma}}$),
$v^{k}_{n,j,\alpha,\sigma}$ ($v^{k}_{n,j,\beta,\bar{\sigma}}$) are
the Bogoliubov quasiparticle amplitudes on the $j$-th site with
corresponding eigenvalues $E^{k}_{n}$. The self-consistent equations
of the mean fields are
\begin{eqnarray}
n_{i,\beta,\uparrow}=&&\sum_{k,n}|u^{k}_{n,i,\beta,\uparrow}|^{2}f(E^{k}_{n}) \nonumber\\
n_{i,\beta,\downarrow}=&&\sum_{k,n}|v^{k}_{n,i,\beta,\downarrow}|^{2}[1-f(E^{k}_{n})].
\end{eqnarray}
The magnitude of the magnetic order on the $i$-th site and the
induced SC pairing correlation in the BAFM layer are defined as,
\begin{eqnarray}
M(i)=&&\frac{1}{2}\sum_{\beta}(n_{i,\beta,\uparrow}-n_{i,\beta,\downarrow})
\nonumber\\
\Delta^{A}_{ij,\alpha\beta}=&&\frac{1}{4}\sum_{k,n}{'}(u^{k}_{n,i,\alpha,\sigma}
v^{k\ast}_{n,j,\beta,\bar{\sigma}}e^{-i\textbf{k}\cdot(\textbf{r}_{j}-\textbf{r}_{i})}
\nonumber\\
&&+v^{k\ast}_{n,i,\alpha,\bar{\sigma}}u^{k}_{n,j,\beta,\sigma}e^{i\textbf{k}\cdot(\textbf{r}_{j}-\textbf{r}_{i})})
\tanh(\frac{E^{k}_{n}}{2k_{B}T}).
\end{eqnarray}

In the calculations, we choose the hopping integrals as
follows:~\cite{wgyin1} Along the $y$ direction, the
$d_{xz}-d_{xz}$ NN hopping integral $t_{1}=0.4$ eV and the
$d_{yz}-d_{yz}$ NN hopping integral $t_{2}=0.13$ eV; they are
exchanged to the $x$ direction; the NN interorbital hoppings are
zero; the NNN intraorbital hopping integral $t_{3}=-0.25$ eV for
both $d_{xz}$ and $d_{yz}$ orbitals, and the NNN interorbital
hopping is $t_{4}=0.07$ eV. The hopping integral $t_{1}$ is taken as
the energy unit. We keep
$J_{1}:J'_{1}:J_{2}:J'_{2}=-4:-1:1:2$.~\cite{CCao1,YZYou1} The
doping level is given by $\delta=n-2.0$.

\section{results}
\begin{figure}[htb]
\begin{center}
\vspace{-.2cm}
\includegraphics[width=230pt,height=150pt]{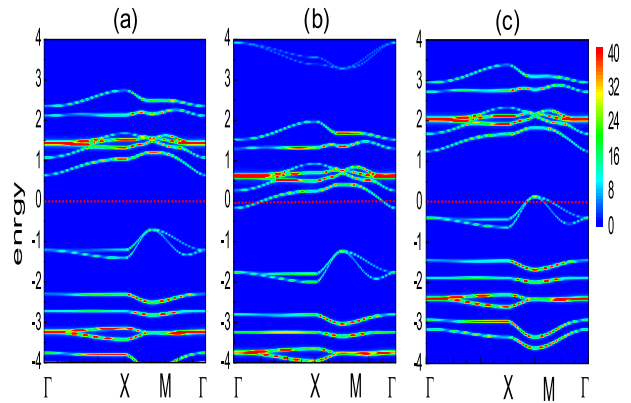}
\caption{(color online) Electronic band structures of $H_{0}$ given
by Eq. (2). The parameters are given at the end of section II of the
text, and $J_{1}=2.0$. (a): at half filling or $n=2.0$; (b): at
electron doping $n=2.1$; and (c): at hole doping $n=1.9$. The color
scale indicates the relative spectra weight.}\label{fig2}
\end{center}
\end{figure}

To begin with, we present the energy band structure at half filling
with $n=2$ in Fig.~\ref{fig2}(a), where $J_{1}=2.0$ is so chosen to
get a band gap $\sim500$meV being in agreement with the first
principle calculations.~\cite{CCao1,XWYan2} For the electron doping
with $n=2.1$, the Fermi level crosses an energy band around the
center of the Brillouin zone [$\Gamma$ point in Fig.~\ref{fig2}(b)],
while it intersects with an energy band around the zone corner at
the hole doping with $n=1.9$ [$M$ point in Fig.~\ref{fig2}(c)].
Although a simple two-orbital model is adopted here, both the
electron and hole doping cases with $\delta=0.1$ are qualitatively
consistent with the first principle calculations.~\cite{XWYan2} In
the presence of the ordered vacancies and BAFM order, the original
two-band structures are splitting to sixteen subbands as a result of
the enlarged unit cell with $8$ sites. At half filling, $8$ lower
bands are occupied, i.e., $1/4$ electron per one subband, while
another $8$ bands above the Fermi energy are unoccupied, resulting
in a band gap in Fig.~\ref{fig2}. For the electron and hole doping
with $\delta=0.1$, the chemical potential crosses one subband which
produce the characteristic features of the Fermi surface and the
metallic BAFM state.

\begin{figure}[htb]
\begin{center}
\vspace{.3cm}
\includegraphics[width=210pt,height=170pt]{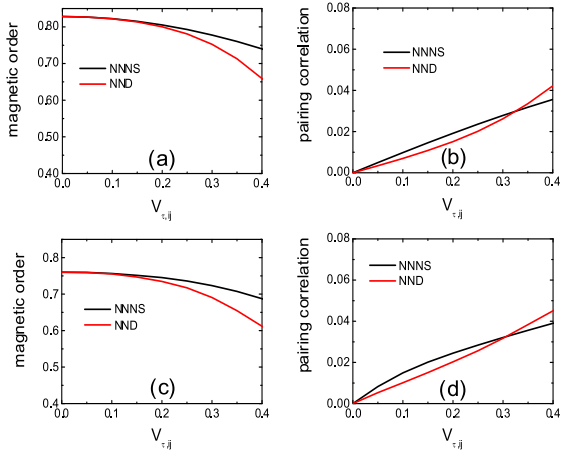}
\caption{(color online) Block AFM moment and the SC pairing
correlation as functions of the effective tunneling strength
$V_{\tau,ij}$. Black curves are for next nearest neighbor
$s_{\pm}$-wave pairing, and red for nearest neighbor $d$-wave
pairing. Upper panel (a) and (b): $n=2.0$ and lower panel (c) and
(d): $n=2.1$.}\label{fig3}
\end{center}
\end{figure}

Motivated by the agreement of the self-consistent mean-field
solutions with the mentioned first principle calculations, we
consider now the proximity effect in BAFM layer induced by the SC in
SC layer. For explicit reason, we choose two possible singlet
pairing symmetries in the SC layer, i.e., the NNN $s_{\pm}$-wave and
the NN $d$-wave symmetries with their respective gap functions
$\Delta_{s_{\pm}}=\Delta_{0}\cos(k_{x})\cos(k_{y})$ and
$\Delta_{d}=\Delta_{0}[\cos(k_{x})-\cos(k_{y})]$, where the former
results in the NNN bond and the latter the NN bond couplings in the
BAFM layer. The interlayer hopping constant $t_{\tau}$ is assumed to
be site independent. Fig.~\ref{fig3} displays the moment of the BAFM
order as a function of the effective tunneling strength
$V_{\tau,ij}$. At the half filling, both symmetries of the SC order
in the SC layer introduce the decrease of the BAFM order and
simultaneously induce the SC correlation with the same symmetries in
the BAFM layer as the tunneling strength increases. A main
difference between the $s_{\pm}$- and the $d$-wave symmetries is the
more pronounced proximity effect in reducing the moment of the BAFM
order produced by the $d$-wave symmetry as the tunneling strength
increase, as shown in Figs.~\ref{fig3}(a) and~\ref{fig3}(b). In the
case of electron doping with $n=2.1$, where the metallic BAFM state
results, although the proximity effect is more pronounced, the
magnetic and the induced SC correlation remain the qualitatively
unchanged, due possibly to the very low total carrier concentration.

\begin{figure}[htb]
\begin{center}
\vspace{-.2cm}
\includegraphics[width=230pt,height=100pt]{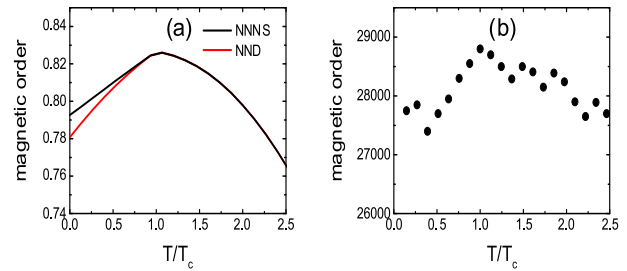}
\caption{(color online) (a) Temperature dependence of the block AFM
moment at $n=2.0$. Black and red curves are for next nearest
neighbor $s_{\pm}$-wave and nearest neighbor $d$-wave pair
couplings, respectively. (b): the re-plotted curve of the neutron
data from Ref.~\onlinecite{WBao1}.}\label{fig4}
\end{center}
\end{figure}

The unique feature of the the effective tunneling in the second
order is it's temperature dependence via the SC pairing
$\Delta_{ij,\alpha\alpha',\sigma,\sigma'}$, which differs from that
in one particle tunneling process.~\cite{mmori1,jxzhu1,andersen1}
The temperature dependence of the SC pairing parameter is modeled by
a phenomenological form with $\Delta=\Delta_{0}\sqrt{1-T/T_{c}}$. We
present the temperature dependence of the magnetic moment in
Fig.~\ref{fig4}(a) for the typical choice of the coupling constants
$g_{\tau}=2V_{\tau,ij}\Delta_{0}=0.25$. As temperature decrease, the
magnetic order increases when temperature is above $T_{c}$, while it
decreases when temperature is below $T_{c}$, resulting in a broad
peak around $T_{c}$. We note that the temperature dependence of the
AFM moment is reminiscent of the neutron diffraction and the
two-magnon experiments [Fig.~\ref{fig4}(b)].~\cite{WBao1,AMZhang}
There is another scenario that the competition between the AFM and
the SC orders in the microscopic coexistence of them may also
produce the decrease of the AFM moment below $T_{c}$. The study of
such possibility is currently under way and the results will be
published elsewhere. It is worthwhile to notice that the sizable
proximity effect relies on the substantial interlayer hoping
constant $t_{\tau}$. Based on the first principle calculation, the
interlayer hopping $t_{\tau}$ was estimated to have a comparable
magnitude with $t_{1}$ possibly due to the high values of electron
mobility from the intercalated alkaline atoms,~\cite{CCao2} and
leads to the highly three dimensional Fermi
surface.~\cite{CCao1,XWYan2}

\vspace*{-.2cm}

\section {SUMMARY AND DISCUSSIONS}

In summary, we have proposed that various seemingly conflict
experiments on the phase separation or coexistence of
superconductivity and BAFM may be explained within a phase
separation scenario by taking into account of the proximity effect
of superconductivity to the neighboring layer of BAFM. We have
theoretically studied the proximity effect to a BAFM layer induced
by adjacent SC layers in a simplified two-orbital model for
Fe-selenides. The proximity effect in reducing the moment of the
BAFM order highly depends on the coupling constant $V_{\tau, ij}$.
For realistic parameters of the interlayer tunneling, our
calculation shows that the superconductivity proximity effect may
result in substantial suppression of the magnetic moment. This is in
contrary to that in the cuprate superconductor, where the coupling
constant $V_{\tau, ij}$ is very small because of small c-axis
hopping integral due to the large anisotropy, and because of the
renormalization of $V_{\tau,ij}$ by a factor proportional to hole
concentrations due to the no double-occupation
condition.~\cite{maekawa} In iron-based superconductor, the
anisotropy of iron-based material is much smaller than in the
cuprate, which lead to a relative larger $t_{\tau}$. And the
moderate correlation effect in iron-based superconductor leads to a
moderate renormalization factors.  As a consequence, the coupling
constant $V_{\tau,ij}$ in iron chalcogenide superconductor should be
moderate.

We remark that we'd be careful in drawing a concrete conclusion to
compare with the experiments. The approximation that only $d_{xz}$
and $d_{yz}$ orbitals are important in the bands close to Fermi
energy is good in terms of the band structures.~\cite{CCao1,XWYan2}
But the maximum magnetic moment in two-orbital model is only
$2\mu$B, smaller than the moment of $3.31\mu$B measured in
experiments.~\cite{ZShermadini,WBao1} The other effect is that we
only calculated the suppression of the BAFM order of the surface
layer of the BAFM domain. According to TEM experiment,~\cite{ZWang1}
each BAFM domain has around ten layers. And the suppression of BAFM
order of the layers in the middle of domain may be more complicated.
In brief, the suppression of the BAFM moment is sizable because of
the moderate coupling constant $V_{\tau,ij}$, and our calculation
may be viewed as a semi-quantitative result.

We also investigated proximity effect for various pairing symmetry
of the SC phase.  It has shown that the SC pairing with NN $d$-wave
symmetry resulted a more pronounced proximity effect in reducing the
moment of the BAFM order than the NNN $s_{\pm}$-wave pairing. The
second order process induced proximity effect has a temperature
dependent as the SC pairing, which may be relevant to the
experimental observations. More remarkable proximity effect was
found in the BAFM state by comparison with the conventional AFM
state, which was the consequence of the frustrated structure and the
associated anisotropic exchange interactions.

\section{acknowledgement}
\par We thank W. Bao, G. Aeppli, Y. Zhou, and T. M. Rice for helpful discussions. This work is supported in part
by Hong Kong's RGC GRF HKU706809 and HKUST3/CRF/09. HMJ is grateful
to the NSFC (Grant No. 10904062), Hangzhou Normal University
(HSKQ0043, HNUEYT).

\section{appendix}
\begin{figure}[htb]
\begin{center}
\vspace{.5cm}
\includegraphics[width=230pt,height=190pt]{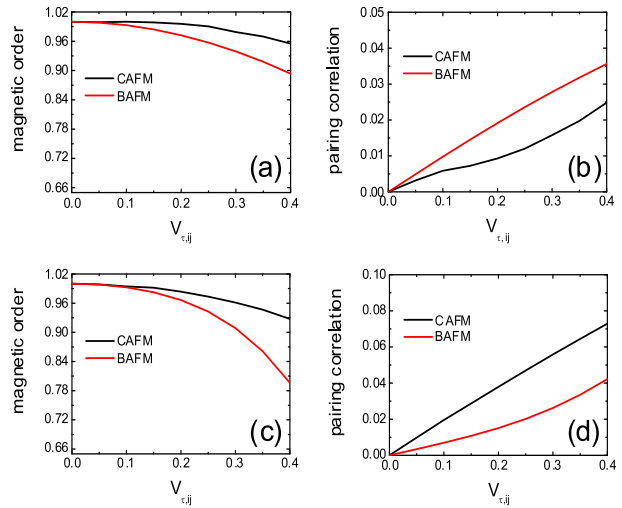}
\caption{(color online) Comparison of the proximity effect between
the block and conventional AFM states. Left column shows the moment
of the AFM order, and right column the SC pairing correlation as
functions of the effective tunneling strength $V_{\tau,ij}$. Upper
panel: for the next nearest neighbor $s_{\pm}$-wave pairing and
lower panel for the nearest neighbor $d$-wave pairing.}\label{fig5}
\end{center}
\end{figure}

In the following, we compare the above proximity effect with that in
the single band conventional AFM (CAFM) system. In order to make the
comparison more convincing, we choose the dispersion
$\varepsilon_{k}=-2t[\cos(k_{x})+\cos(k_{y})]-4t'\cos(k_{x})\cos(k_{y})-\mu$
with $t=t_{1}$ and $t'=t_{3}$, which gives rise to the similar
energy band width with that in the above two-orbital model and is
close to the case of the cuprates. The AFM order is introduced by
the AFM exchange interaction $J\sum_{\langle
ij\rangle}\textbf{S}_{i}\cdot\textbf{S}_{j}$ between the NN sites.
At the half filling $n=1$, we find that $J=1.6$ produces the
comparable band gap and the electron polarization as in the above
BAFM state. In Fig.~\ref{fig5}, we present the magnitude of the
magnetic order and the induced pairing correlation as a function of
the effective tunneling $V_{\tau,ij}$. The upper panel shows the
results for the NNN $s_{\pm}$-wave pairing and the lower panel the
results for the NN $d$-wave pairing. In the figure, the magnitude of
the magnetic order in both cases is renormalized. The proximity
effect in reducing the AFM order is more pronounced for the BAFM
state as shown in Figs.~\ref{fig5}(a) and~\ref{fig5}(c). As for the
induced pairing correlation, the larger correlation is found in the
BAFM state for the $s_{\pm}$-wave paring and in the CAFM state for
the $d$-wave pairing, as displayed in Figs.~\ref{fig5}(b)
and~\ref{fig5}(d), respectively.

\vspace*{.2cm}
\begin{figure}[htb]
\begin{center}
\vspace{.3cm}
\includegraphics[width=230pt,height=120pt]{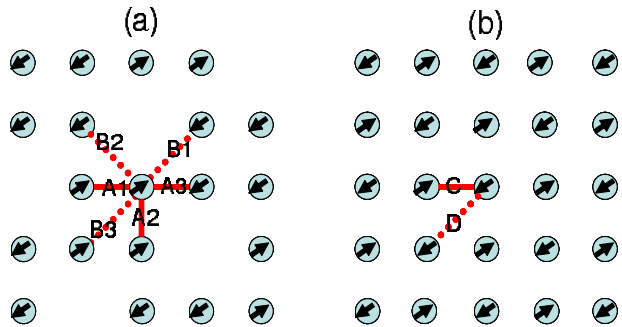}
\caption{(color online) Comparison of the spin structures and their
respective NN and NNN bonds. (a): block AFM state; (b): conventional
AFM state.}\label{fig6}
\end{center}
\end{figure}
\vspace*{.0cm}

We can understand the above results by considering the different
spin configurations of the BAFM and CAFM orders, as shown in
Fig.~\ref{fig6}. In the BAFM state, when two electrons transfer from
the BAFM layer to the SC one, the energy changes due to the bonds
breaking for the NN bond coupling are $\Delta
E^{\uparrow\uparrow}_{NN}=|J_{1}|$ [A1 and A2 bonds in
Fig.~\ref{fig6}(a)] and $\Delta
E^{\uparrow\downarrow}_{NN}=7|J_{1}|/4$ [A3 bond in
Fig.~\ref{fig6}(a)] for the electron pairs with ferromagnetic and
antiferromagnetic alignments, respectively. On the other hand, the
energy changes due the bonds breaking for the NNN bond are $\Delta
E^{\uparrow\uparrow}_{NNN}=9|J_{1}|/4$ [B3 bond in
Fig.~\ref{fig6}(a)] and $\Delta
E^{\uparrow\downarrow}_{NNN}=5|J_{1}|/2$ [B1 and B2 bonds in
Fig.~\ref{fig6}(a)]. As a result, the proximity effect in reducing
the moment of the AFM order by the $d$-wave pairing is more
remarkable than that by the $s_{\pm}$ one. However, the energy
changes due to the bonds breaking in the CAFM state are $\Delta
E_{NN}=7|J|$ [C bond in Fig.~\ref{fig6}(b)] and $\Delta
E_{NNN}=8|J|$ [D bond in Fig.~\ref{fig6}(b)] for the NN and NNN band
couplings. Therefore, the proximity effect in reducing the moment of
the AFM in the CAFM state is rather weak for both the $s_{\pm}$- and
$d$-wave pairing couplings. As for the induced pairing correlation,
the extent of the match between the AFM order configuration and the
singlet SC pairing largely determines the magnitude of the induced
pairing correlation. For example, the CAFM matches well with the NN
$d$-wave pairing, so that one can expect a large induced pairing
correlation without the severe decrease of the AFM order as
displayed in Figs.~\ref{fig5}(c) and~\ref{fig5}(d).


\begin{thebibliography}{60}
\bibitem{JGuo} J. Guo, S. Jin, G. Wang, S. Wang, K. Zhu, T. Zhou, M. He, and X. Chen, Phys. Rev. B \textbf{82},
        180520(R) (2010).
\bibitem{AKrzton} A. Krzton-Maziopa, Z. Shermadini, E. Pomjakushina, V. Pomjakushin, M. Bendele, A. Amato,
        R. Khasanov, H. Luetkens, and K. Conder, J. Phys.: Condens. Matter \textbf{23}, 052203 (2011).
\bibitem{MHFang} M. H. Fang, H. D. Wang, C. H. Dong, Z. J. Li, C. M. Feng, J. Chen, and H. Q. Yuan, Europhys. Lett.
        \textbf{94}, 27009 (2011).
\bibitem{YZhang1}Y. Zhang, L. X. Yang, M. Xu, Z. R. Ye, F. Chen, C. He, H. C. Xu, J. Jiang, B. P. Xie, J. J. Ying,
        X. F. Wang, X. H. Chen, J. P. Hu, M. Matsunami, S. Kimura, and D. L. Feng, Nature Materials \textbf{10}, 273
       (2011).
\bibitem{Cruz1} C. de la Cruz, Q. Huang, J. W. Lynn, J. Li, W. Ratcliff II, J. L. Zarestky, H. A. Mook, G. F. Chen,
        J. L. Luo, N. L. Wang, and P. Dai, Nature \textbf{453}, 899 (2008).
\bibitem{FMa1} F. Ma, Z. Y. Lu, and T. Xiang, Phys. Rev. B \textbf{78}, 224517 (2008);
        Front. Phys. China \textbf{5}, 150 (2010).
\bibitem{XWYan1} X. W. Yan, M. Gao, Z. Y. Lu, and T. Xiang, Phys. Rev. Lett. \textbf{106}, 087005 (2011).
\bibitem{FMa2} F. Ma, W. Ji, J. Hu, Z. Y. Lu, and T. Xiang, Phys. Rev. Lett. \textbf{102}, 177003 (2009).
\bibitem{WBao2} W. Bao, Y. Qiu, Q. Huang, M. A. Green, P. Zajdel, M. R. Fitzsimmons, M. Zhernenkov, S. Chang,
        M. Fang, B. Qian, E. K. Vehstedt, J. Yang, H. M. Pham, L. Spinu, and Z. Q. Mao,
        Phys. Rev. Lett. \textbf{102}, 247001 (2009).
\bibitem{SLi1} S. Li, C. de la Cruz, Q. Huang, Y. Chen, J. W. Lynn, J. Hu, Y. L. Huang, F. C. Hsu, K. W. Yeh,
        M. K. Wu, and P. Dai, Phys. Rev. B \textbf{79}, 054503 (2009).
\bibitem{WBao1} W. Bao, Q. Huang, G. F. Chen,M. A. Green, D. M. Wang, J. B. He, X. Q. Wang, and Y. Qiu,
        Chinese Phys. Lett. \textbf{28}, 086104 (2011).
\bibitem{ZShermadini} Z. Shermadini, A. Krzton-Maziopa, M. Bendele, R. Khasanov, H. Luetkens, K. Conder, E. Pomjakushina,
        S. Weyeneth, V. Pomjakushin, O. Bossen, and A. Amato, Phys. Rev. Lett. \textbf{106}, 117602 (2011).
\bibitem{yzhou} Y. Zhou, D.-H. Xu, F.-C. Zhang, and W.-Q. Chen, Europhys. Lett. \textbf{95}, 17003 (2011).
\bibitem{RYu} R. Yu, J.-X. Zhu, and Q. Si, Phys. Rev. Lett. \textbf{106}, 186401 (2011).
\bibitem{RHYuan} R. H. Yuan, T. Dong, G. F. Chen, J. B. He, D. M. Wang, and N. L. Wang,
        e-print arXiv:1102.1381 (to be published).
\bibitem{AMZhang} A. M. Zhang, J. H. Xiao, Y. S. Li, J. B. He, D. M. Wang, G. F. Chen, B. Normand, Q. M. Zhang,
        and T. Xiang, e-print arXiv:1106.2706 (unpublished).
\bibitem{Shermadini1} Z. Shermadini, A. Krzton-Maziopa, M. Bendele, R. Khasanov, H. Luetkens, K. Conder, E. Pomjakushina,
        S. Weyeneth, V. Pomjakushin, O. Bossen, and A. Amato, Phys. Rev. Lett. \textbf{106}, 117602 (2011).
\bibitem{FChen1} F. Chen, M. Xu, Q. Q. Ge, Y. Zhang, Z. R. Ye, L. X. Yang, Juan Jiang, B. P. Xie, R. C. Che, M. Zhang,
        A. F. Wang, X. H. Chen, D. W. Shen, X. M. Xie, M. H. Jiang, J. P. Hu, D. L.
        Feng, e-print arXiv:1106.3026 (unpublished).
\bibitem{Torchetti1} D. A. Torchetti, M. Fu, D. C. Christensen, K. J. Nelson, T. Imai, H. C. Lei, and C. Petrovic,
        Phys. Rev. B \textbf{83}, 104508 (2011).
\bibitem{ZWang1} Z. Wang, Y. J. Song, H. L. Shi, Z. W. Wang, Z. Chen, H. F. Tian, G. F. Chen, J. G. Guo, H. X. Yang,
        and J. Q. Li, Phys. Rev. B \textbf{83}, 140505(R) (2011).
\bibitem{wli1} W. Li, H. Ding, P. Deng, K. Chang, C. Song, K. He, L. Wang, X. Ma, J.-P. Hu, X. Chen, and Q.-K. Xue,
       e-print arXiv:1108.0069 (unpublished).
\bibitem{aeppli} G. Aeppli, D. Bishop, C. Broholm, E. Bucher, K. Siemensmeyer,  M. Steiner, and N. Stusser, Phys. Rev. Lett. \textbf{63}, 676 (1989).
\bibitem{varma} E. I. Blount, C. M. Varma, G. Aeppli, Phys. Rev. Letts. \textbf{64}, 3074 (1990).
\bibitem{mcmillan1} W. L. McMillan, Phys. Rev. \textbf{175}, 537 (1968).
\bibitem{wgyin1} W.-G. Yin, C.-C. Lee, and W. Ku, Phys. Rev. Letts. \textbf{105}, 107004 (2010).
\bibitem{CCao1} C. Cao, and J. Dai, Phys. Rev. Lett. \textbf{107}, 056401 (2011).
\bibitem{YZYou1} Y.-Z. You, H. Yao, and D.-H. Lee, Phys. Rev. B \textbf{84}, 020406(R) (2011).
\bibitem{XWYan2} X.-W. Yan, M. Gao, Z.-Y. Lu, and T. Xiang, Phys. Rev. B \textbf{83}, 233205 (2011).
\bibitem{mmori1} M. Mori, and S. Maekawa, Phys. Rev. Lett. \textbf{94}, 137003 (2005).
\bibitem{jxzhu1} J.-X. Zhu, and C. S. Ting, Phys. Rev. B \textbf{61}, 1456 (2000).
\bibitem{andersen1} B. M. Andersen, I. V. Bobkova, P. J. Hirschfeld, and Yu. S. Barash, Phys. Rev. B \textbf{72}, 184510
  (2005).
\bibitem{CCao2} C. Cao, private communications.
\bibitem{maekawa} M. Mori, T. Tohyama and S. Maekawa, J. Phys. Soc. Jpn. \textbf{75}, 034708 (2006)
\end{thebibliography}
\end{document}